\documentclass[apj]{emulateapj}

\shorttitle{Hydrocarbon Core in IRAS 04296 and IRAS 05341}
\shortauthors{Goto et al.}

\begin{document}

\title{
Diffraction-limited 3\micron~Spectroscopy of IRAS 04296+3429
and IRAS 05341+0852: \\ Spatial Extent of Hydrocarbon Dust Emission
and Dust Evolutionary Sequence\altaffilmark{1}}

\author{Miwa Goto,\altaffilmark{2,3} Sun Kwok,\altaffilmark{4} Hideki
Takami,\altaffilmark{5} Masa Hayashi,\altaffilmark{5}
W. Gaessler,\altaffilmark{2} Yutaka Hayano,\altaffilmark{5} Masanori
Iye,\altaffilmark{6} Yukiko Kamata,\altaffilmark{6} Tomio
Kanzawa,\altaffilmark{5} Naoto Kobayashi,\altaffilmark{7} Yosuke
Minowa,\altaffilmark{7} Ko Nedachi,\altaffilmark{5} Shin
Oya,\altaffilmark{5} T.-S. Pyo,\altaffilmark{5}
D. Saint-Jacques,\altaffilmark{8} Naruhisa Takato,\altaffilmark{5}
Hiroshi Terada,\altaffilmark{5} \protect\and Th.
Henning\altaffilmark{2}}

\email{mgoto@mpia.de}

\altaffiltext{1}{Based on data collected at Subaru Telescope,
operated by the National Astronomical Observatory of Japan}

\altaffiltext{2}{Max-Planck-Institut f\"ur Astronomie, K\"onigstuhl 17,
Heidelberg D-69117, Germany}

\altaffiltext{3}{Visiting astronomer at the Institute for
Astronomy, University of Hawaii}


\altaffiltext{4}{Department of Physics, University of Hong Kong, Hong
Kong, China}

\altaffiltext{5}{Subaru Telescope, 650 North A`ohoku Place, Hilo,
HI 96720}

\altaffiltext{6}{National Astronomical Observatory of Japan, Osawa,
Mitaka, Tokyo 181-8588, Japan}

\altaffiltext{7}{The University of Tokyo, Osawa, Mitaka, Tokyo 181-0015,
Japan}

\altaffiltext{8}{Groupe d'astrophysique, Universit\'e de Montr\'eal,
2900 Boul. \'Edouard-Monpetit, Montr\'eal (QC) H3C 3J7, Canada}

\begin{abstract}

We present 3~$\mu$m spectroscopy of the carbon-rich proto-planetary
nebulae IRAS 04296+3429 and IRAS 05341+0852 conducted with the
adaptive optics system at the Subaru Telescope. We utilize the nearly
diffraction-limited spectroscopy to probe the spatial extent of the
hydrocarbon dust emitting zone. We find a hydrocarbon emission core
extending up to 100--160~mas from the center of IRAS 04296+3429,
corresponding to a physical diameter of 400--640~AU, assuming a
distance of 4~kpc. On the other hand, we find that IRAS 05341+0852 is
not spatially resolved with this instrumentation. The physical extent
of these proto-planetary nebulae, along with the reanalyzed data of
IRAS 22272+5435 published previously, suggests a correlation between
the physical extent of the hydrocarbon dust emission and the spectral
evolution of the aliphatic to aromatic features in these post-AGB
stars. These measurements represent the first direct test of the
proposed chemical synthesis route of carbonaceous dust in the
circumstellar environment of evolved stars.

\end{abstract}

\keywords{stars: AGB and post-AGB --- circumstellar matter --- stars:
individual (IRAS 04296+3429, IRAS 05341+0852, IRAS 22272+5435) ---
dust, extinction --- ISM: evolution --- infrared: ISM}

\section{Introduction}

The origin of carbonaceous grains observed in the interstellar medium
and in the early solar system is a topic of great interest. These
grains are composed of a chain or a network of carbon atoms, bonded to
hydrogen atoms at their periphery with physical and chemical
properties intermediate between those of simple molecules and of bulk
solid particles. Due to the versatility of carbon atoms in creating
various hybridized bonds, these carbonaceous grains can have very
complex chemical structures \citep{hen98,ehr00,pen02,hen03}.

Recent infrared spectroscopic observations from space have revealed
that carbon-rich post-asymptotic giant branch (post-AGB) stars undergo
rapid synthesis of hydrocarbon grains with both aromatic and aliphatic
signatures. The ``aromatic'' hydrocarbon is characterized by
benzene-like rings in which carbon atoms are connected to each other
with conjugated double bonds, while the ``aliphatic'' hydrocarbon
consists of linear chains or a network of carbon atoms connected by
saturated bonds. The production and ejection of these grains may play
a significant role in the chemical enrichment of the interstellar
medium and the early solar system \citep{kwo04}. Specifically, while
carbon-rich AGB stars show either the 11.3 $\mu$m feature due to SiC
or a featureless continuum due to amorphous carbon
\citep{hen01,cle03}, aromatic features at 3.3, 6.2, 7.7, 8.6, and 11.3
$\mu$m and aliphatic features at 3.4 and 6.9~$\mu$m begin to appear in
proto-planetary nebulae (PPN, Kwok, Volk, \& Hrivnak 1999). We now
know that the aliphatic to aromatic transformation of the
circumstellar dust from PPN to planetary nebulae coincides with
environmental changes of increasing stellar temperature, less
protective dusty environment, and the outbreak of harsh UV radiation
\citep{kwo01}. In order to test any photochemical models, we first
need a good knowledge of the spatial distribution of the emitting
materials. The observations of these objects in the short transition
phase between AGB stars and planetary nebulae are therefore of primary
importance in our understanding of the chemical synthesis of
carbonaceous grains.

IRAS 04296+3429 along with IRAS 05341+0852 and IRAS 22272+5435 belong
to a small group of PPNs with extreme carbon-rich chemistry, first
recognized by the presence of an unidentified 21~$\mu$m emission
feature in their IRAS/LRS spectra \citep{kwo89,khg95}. The
observational properties of these sources, taken from the literature,
are summarized in Table~\ref{tb1}. This class of PPN is most
exclusively characterized by prominent emission features of highly
aliphatic material with broad emission at 3.4~$\mu$m alongside the
common aromatic feature at 3.3~$\mu$m first detected by \citet{geb90}
and \citet{geb92}, subsequently classified in \citet{geb97}. Since the
hydrocarbon dust is synthesized in the circumstellar environment, if
it undergoes temporal modification it will leave a record of unique
spatial distribution in the circumstellar envelope. With sufficient
spatial resolution, we may observe the history of dust formation as a
change of hydrocarbon dust structure with distance from the central
star. Such observations provide a critical test for the proposed
evolutionary schemes \citep[e.g., ][]{kwo99}. However, PPNs are
usually small in angular extent and require sub-arcsecond resolution
to image. It is only recently that the imaging resolution in the
infrared has become good enough to perform such experiments
\citep{got03}. In this paper, we present high-angular resolution
imaging spectroscopy of the distribution of circumstellar hydrocarbon
dust using an adaptive optics (AO) system.

\begin{figure}
\includegraphics[angle=-90,width=0.48\textwidth]{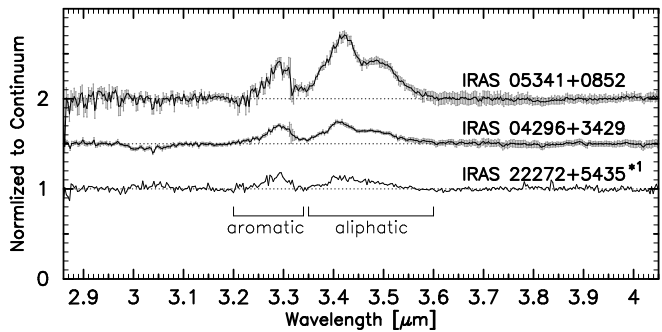}
\caption{The spectra of IRAS 05341+0852 and IRAS 04296+3429 after
division by a polynomial fit to the continuum. Also plotted is the
spectrum of IRAS 22272+5435 from \citet{got03}. These three PPNs are
extremely carbon-rich with a specific aliphatic emission features at
3.4~$\mu$m along with more common aromatic emission at 3.3~$\mu$m. The
intensity of the aliphatic emission relative to the aromatic feature
decreases with the physical development of the nebulae. The ratio of
the whole emission feature to the continuum emission exhibits the same
trend. Note that the spectrum of IRAS 22272+5435 is extracted from the
nebula, but in a region offset from the central star, in order to
highlight the hydrocarbon emission features. The feature to continuum
ratio is much smaller when the central star is included, which makes
the trend above even stronger.\label{f1}}
\end{figure}

\section{Observations}

The 3~$\mu$m spectra of IRAS 04296+3429 and IRAS 05341+0852 were
obtained at the Subaru Telescope atop Mauna Kea on 2001 December 23
with the Infrared Camera and Spectrograph (IRCS; Tokunaga et al. 1998;
Kobayashi et al. 2000). An adaptive optics system was used as a
front-end instrument to feed wavefront-compensated images to IRCS. The
Subaru AO is a 36-element curvature-based system installed at the
telescope Cassegrain port \citep{gae02,tak04}. It delivers
diffraction-limited images at wavelengths longer than 2~$\mu$m. A
medium-resolution grism was used with a 0\farcs30 slit to provide
spectra with a resolving power of $\lambda / \Delta \lambda =$
600--800. The plate scale of the spectrograph was 58~mas/pixel along
the slit. The slit was aligned to PA $=$ 28\arcdeg~and 98\arcdeg~for
IRAS 04296+3429, along the disk and the bipolar lobe projected on the
sky \citep{sah99}. The slit was rotated to PA $=$ 45\arcdeg~for IRAS
05341+0852 which is along the elongation of the nebula at optical
wavelengths \citep{uet00}. The visible central stars of IRAS
04296+3429 (1200-02336188 in the USNO A-2.0 Catalogue; $R =$ 13.7~mag)
and IRAS 05341+0852 (0975-01778353; $R =$ 12.7~mag) were used as the
wavefront references for the AO system. The spectrograms were recorded
by wobbling the AO system's tip-tilt mirror by $\pm$1\farcs5 along the
slit. A G0.5V star HR~660 ($R =$ 4.5~mag) was observed at the
beginning of the observation through similar airmass to calibrate for
atmospheric transmission. The mismatch in airmass between this
reference and the observed objects was smaller than 0.1. In order to
use HR~660 as the standard star for the point-spread function (PSF) as
well, HR~660 was dimmed by neutral density filters in front of the
wavefront sensor until the output counts of the avalanche photodiodes
were comparable to that of the objects, so that the corrected
wavefront error would be similar. The observation of IRAS 04296+3429
was completed within 1.3 hrs of observing HR~660 for the first slit
position at PA$=$98\arcdeg.
HR~660 may not be as good a PSF standard for IRAS 05341+0852 as this
was observed nearly 4 hours later. The PSF as measured with HR~660
provides an independent measurement to test the spatial extent of IRAS
04296+3429, as discussed in \S3.1. The spectroscopic flat field was
made at the end of the night using a halogen lamp installed at the
calibration unit in front of the instrument entrance window. The
seeing was good at the beginning of the night (0\farcs55 at $R$),
deteriorating towards the end of the observation (0\farcs86 at
$R$). The details of the observation of IRAS 22272+5435 are presented
elsewhere \citep{got03}.

\section{Data analysis}

\subsection{Non-linearity correction}

Data reduction begins with correcting each frame for detector
non-linearity, since even a slight saturation of the detector response
affects the PSF measurement so crucial to the present study. Suppression
of pixel counts at the peak of the PSF modifies the spatial profile by 
slightly 
flattening it. Therefore, the measured spatial width is
increased. Special care must be taken to account for the pixel
counts in the pedestal readout. A typical readout of an infrared
detector array consists of three steps: a reset of the entire array,
after which integration starts immediately; a first readout to sample
the pixel counts at the beginning of the exposure; and a second
readout after the desired integration time. Then the pedestal count
from the first readout is subtracted, yielding the net signal. This
net signal is usually monitored during observations not to exceed the
detector linearity limit.

However, for observations in high-flux regime such as the ones done
with the AO system, the integration time can be so short that it is
comparable to the detector readout time. Substantial photon flux is
already accumulated in the pixel before the first readout. The
pedestal counts can potentially push the second readout up into the
non-linear regime, a situation which cannot be recognized when only
the net signal is monitored. The pedestal count depends on how soon
the relevant pixel is read after the array reset, and therefore
depends on the location of the pixel in the array. The magnitude of
the non-linearity correction, including the position-dependent
pedestal count, amounts to not more than 8\% in most of the
observations with our setup using a pixel readout of 0.41~s for the
entire array, an integration time of 8~s, and with maximum net counts
restricted to values below 5,000~ADU.



\subsection{Spectroscopy}

Following the above described non-linearity correction, consecutive
spectrograms were subtracted from each other to eliminate background
sky emission. Pixel-to-pixel detector response variation was
normalized by ratioing with dark-subtracted flat-field images. The
extraction aperture was defined at a fixed distance from the profile
centroid, and the enclosed pixel counts were summed up to produce a
one-dimensional spectrum. Wavelength calibration was performed by
maximizing the correlation of resulting spectrum with a model
atmospheric transmission curve computed by ATRAN \citep{lor92}.
Telluric atmospheric absorption features were removed by dividing by
the spectrum of the spectroscopic standard star HR~660.


The results are shown in Figure~\ref{f1} for IRAS 04296+3429 and IRAS
05341+0852, after normalization by a polynomial function fitted to the
continuum emission in order to highlight the hydrocarbon emission
features at 3.2--3.6~$\mu$m. The spectrum of IRAS 22272+5435 extracted
from the off-center region of the nebula, originally published in
\citet{got03}, is shown for comparison. Two parallel trends are
noticeable. First, the aliphatic emission feature at 3.4~$\mu$m,
specific to this class of carbon-rich PPNs, diminishes as one
progresses from IRAS 05341+0852 to IRAS 04296+3429 and to IRAS
22272+5435, with respect to the aromatic emission at 3.3~$\mu$m. At
the same time, the overall intensity of hydrocarbon emission above the
continuum decreases.

Large-scale ($>$ 0\farcs5) variations of the spectral shape of the
emission feature were checked for IRAS 04296+5435 and IRAS 05341+0852
in the same way as was done for IRAS 22272+5435, by shifting the
extraction aperture from the central star \citep{got03}. The flux at
the off-center regions is dominated by the PSF halo of the central
star, and no significant difference was observed between the spectra
extracted at the central stars and those taken from the off-center
regions.

\begin{figure}
\includegraphics[angle=-90,width=0.48\textwidth]{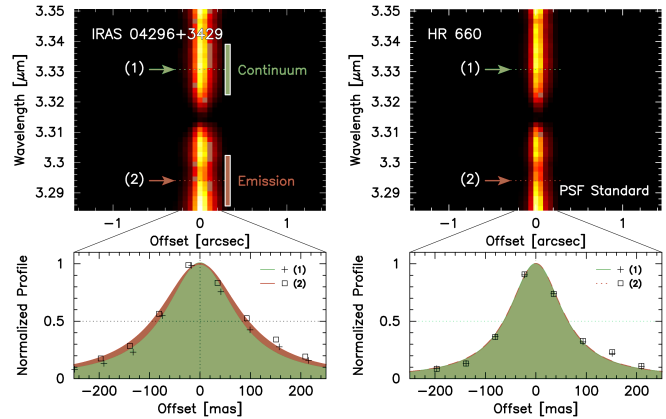}
\caption{Top left panel: A close-up view of the spectrogram of IRAS
04296+3429 at 3.29--3.35~\micron. The spectrogram is pair-subtracted
and flat-fielded. The aromatic hydrocarbon dust produces emission at
3.3~$\mu$m which is marked as ``Emission'' in the spectrogram. The
3.32--3.34~\micron~region, in between the 3.3~$\mu$m aromatic and the
3.4~$\mu$m aliphatic feature, is labeled as ``Continuum''. Bottom
left: A one-dimensional slice of the spectrogram cut along the slit at
the emission wavelength (marked by rectangles) and on the continuum
(crosses). The green and red filled curves are Lorentzian functions
used to fit the pixel data at the continuum and at the emission
feature, respectively. It is evident that IRAS 04296+3429 is larger at
the 3.3~$\mu$m aromatic feature than at the continuum wavelength.
Right panels: Same as left panels, but for the standard star HR~660.
In this case the spatial profiles of the two spectral regions are
almost identical.\label{f2}}
\end{figure}

\subsection{Spatial profiles}

Adaptive optics provides superb spatial resolution for imaging and
spectroscopy. However, this resolution is highly variable when
operating at the system's performance limit, reflecting the
variability of observing conditions. The reference PSF varies
according to the seeing, the brightness of the wavefront reference
source, and the observing wavelength (therefore the color of each
object). This variable PSF makes the interpretation of the spatial
extent of observed sources a challenge. For instance, if a stellar
image is larger than a known unresolved reference observed previoulsy,
there is no way to tell whether that source has indeed a circumstellar
structure, or if the seeing has simply become worse. This uncertainty
can be eliminated if a point source reference is observed at the same
time with the same instrumental setup, and preferably in the same
frame. This is often difficult to achieve.

We used spectroscopy as a means of sampling a reference PSF within
frames. Spectroscopy amounts to simultaneous imaging at multiple
wavelengths over hundreds of spectral elements, though spatially one
dimensional. If certain spectral features are spatially extended, this
is most sensitively detected by differentiating the spatial profile
from that of the immediate continuum region. The scheme is illustrated
in Figure \ref{f2}. In the present observations, we have compared the
cross section of the spectrogram at the hydrocarbon dust emission
wavelength with the nearby continuum. The spatial profile of IRAS
04296+3429 in Figure 2 is wider at the level of the emission feature
by approximately 16\%, compared to the continuum wavelength. This
difference in the widths of the two spectral regions is not large, but
it is real, otherwise the spatial profiles at these two spectral
regions would be identical, as it is the case for the comparison star
HR~660 shown in the same figure.


Since spatial profiles are sparsely sampled in pixelized data, the
pixel data must be interpolated in order to quantitatively parametrize
the spatial profile. Empirically, we found that the observed PSF is
better approximated by a Lorentzian than by a single component
Gaussian, given the high wings resulting from of the first ring and
the seeing halo around the diffraction core (Fig.~\ref{f2}). The bad
pixels with outlying responses were deliberately left uncorrected,
because smoothing outlier pixels modifies the intrinsic spatial
profile, leading to an incorrect spatial width.
The full-width-at-half-maximum (FWHM) of the spatial profiles was
derived from these fits, and was measured at each spectral element
from 2.9 to 4.1~$\mu$m. The ``FWHM spectrum'', the variation of FWHM
as a function of the wavelength, was measured separately in each frame
to preserve spatial resolution. They were then combined together for
each source. The results are shown in Figure \ref{f3}. The spatial
width of IRAS 04296+3429 is clearly extended near the 3.3--3.4~$\mu$m
hydrocarbon emission features, compared to its size at continuum
wavelengths.

The spectroscopic data for IRAS 22272+5435, obtained by \citet{got03},
were reprocessed in the same manner as described here, and the
resulting FWHM spectrum is shown in Figures \ref{f3} for comparison
with IRAS 04296+3429 and IRAS 05341+0852.


\subsection{Deconvolution}

The spatial width we measured in \S 3.3 is a stellar angular diameter
convolved with the system PSF. In order to derive the intrinsic
dimension of the emitting region, we need to deconvolve the PSF from
the apparent spatial width. For this task, we assume the observed
spatial width is a simple squared sum of the intrinsic angular size
and the size of the PSF at that wavelength,
\[
\theta^2_{\rm obs}(\lambda) = \theta^2_{\rm \ast}(\lambda)
                              + \theta^2_{\rm PSF}(\lambda).
\]
Obtaining the correct PSF, $\theta_{\rm PSF}(\lambda)$, is not a
trivial task, since the wavelength-dependent PSF changes with AO
performance at the time of observing. We calculate the intrinsic
physical dimension $\theta_{\rm \ast}(\lambda)$ in two ways, assuming
different $\theta_{\rm PSF}(\lambda)$ at deconvolution.

\begin{figure}
\includegraphics[angle=-90,width=0.48\textwidth]{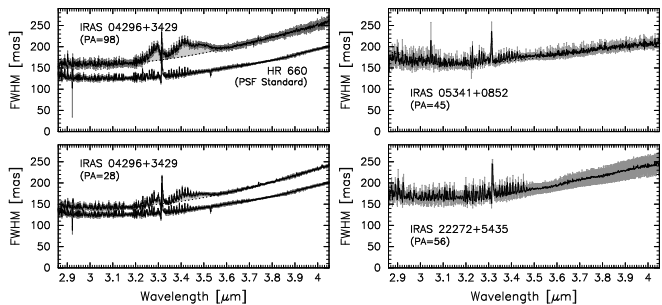}
\caption{Variation of the FWHM of the sources as a function of
wavelength. Top left: the spatial extent of IRAS 04296+3429 at
position angle 98$\degr$ along the bipolar lobe. The smooth increase
in FWHM with wavelength represents the diffraction-limited behavior of
a PSF. This wavelength dependence is fitted by a dotted line, and any
excess above it is emphasized by light gray shade. The error bars are
given by the dispersion of the FWHM in the repeated measurements
(shown in dark gray). The spatial extent of the standard star HR~660
is plotted in the same panel for comparison, as well as the polynomial
fit to the continuum of the FWHM spectrum of IRAS 04296+3429, used as
the point spread functions in the later deconvolution. Bottom left:
variation of spatial extent of IRAS 04296+3429 measured at position
angle 28$\degr$ along the disk. Top right: variation of spatial extent
of IRAS 05341+0852 measured along the position angle 45$\degr$. Bottom
right: variation of spatial extent of IRAS 22272+5435 measured along
the position angle 56$\degr$. \label{f3}}
\end{figure}

\begin{figure}
\includegraphics[angle=-90,width=0.48\textwidth]{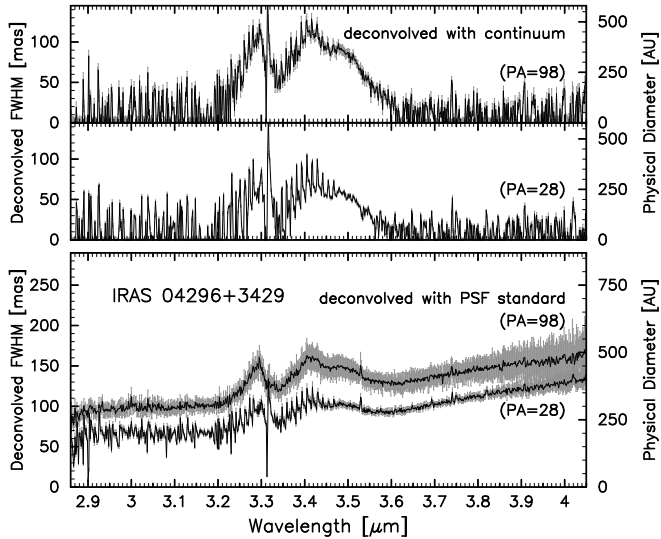}
\caption{Variation of the size of the hydrocarbon emission zone as a
function of wavelength after deconvolution. Top: FWHM spectra
deconvolved with the continuum FWHM of IRAS 04296+3429 itself;
assuming the continuum emission is a point source, and smoothly
increases with wavelength as is in the polynomial fit shown in dotted
lines in Figure \ref{f3}. Bottom: FWHM spectra deconvolved using the
FWHM of the standard star HR~660 as the PSF. The upper trace is for
measurements along PA=98$\degr$, and the lower trace for
PA=28$\degr$.  The angular scale on the left vertical axis is
converted to physical size on the right vertical axis, assuming a
distance of 4~kpc \citep{mei97}.
\label{f4}}
\end{figure}

First, we assume the continuum emission of IRAS 04296+3429 is an
unresolved point source. We measure only the {\it relative} dimension
of the hydrocarbon emitting region with respect to the continuum dust
emission. The apparent diameter of IRAS 04296+3429 smoothly increases
at longer wavelengths except near the hydrocarbon emission features
and at the spikes of heavy absorption by methane in the telluric
atmosphere at 3.2--3.4~$\mu$m. A polynomial function is fit to the
continuum of the FWHM spectrum, and is taken as $\theta_{\rm
PSF}(\lambda)$ (Fig.~\ref{f3}). The large $\theta_{\rm PSF}(\lambda)$
assumed here provides a stringent lower limit for the extent of the
hydrocarbon emitting zone. The apparent angular diameter of the
continuum emitting region is 180~mas at 3.4~\micron~at PA=98$\degr$,
as is shown by the fit in Figure~\ref{f3}. On the other hand, the
apparent FWHM is well above 200~mas at the hydrocarbon emission
features, yielding a minimum physical diameter for the hydrocarbon
emitting region of $\sim$100~mas (Fig. \ref{f4}), corresponding to
400~AU in diameter assuming a distance of 4~kpc for IRAS 04296+3429
\citep{mei97}.

In an alternative approach, we assume the system PSF was stable
throughout the observations and simply use the apparent diameter of
HR~660 as $\theta_{\rm PSF}(\lambda)$ to deconvolve the FWHM of IRAS
04296+3429. This assumption allows an {\it absolute} estimate of the
actual physical dimension of IRAS 04296+3429. The deconvolved FWHM of
IRAS 04296+3429 by this method is 160~mas at 3.4~$\mu$m. The
corresponding physical dimension of the hydrocarbon core is 640~AU (PA
$=$ 98\arcdeg) in diameter (Fig.~\ref{f4}).


IRAS 04296+3429 is apparently larger than HR~660 at all the
wavelengths observed in the present study (Fig.~\ref{f3}), which may
suggest that even the continuum emission is indeed extended and not a
point source as we assumed in the first method. However, it is not
clear which method described above is better suited to this particular
case, as the dust emission tops the stellar component exactly at the
relevant wavelengths near 3--4 ~$\mu$m \citep{kwo93,uet00,fuj02},
which implies the continuum emission is still dominated by the
stellar point source; although there could be a substantial
contribution to the continuum emission in the 3~$\mu$m region from
very small, stochastically heated grains.

\section{Discussion}

 From {\it HST} optical imaging, we know that IRAS 04296+3429 consists
of three major components: a large spherical halo; a disk seen in the
scattered light; and a bipolar lobe nearly vertical to the disk,
plunging into the surrounding halo \citep{sah99}. The halo is the
remnant of the circumstellar material ejected during the previous
asymptotic giant branch phase. The bipolar lobe, delineated by its
interface to the halo, is supposed to be collimated by the disk. This
geometry is typical of PPNs and we assume that they were formed in the
order of the halo, the disk, and the bipolar lobes. The newly found
hydrocarbon emitting region is much smaller in size than all of the
above three components, and does not correspond to any of the
structures previously imaged in visible light. The double-peaked
spectral energy distribution observed in this object suggests that the
dust shell is detached from the central star and large-scale mass loss
has ceased some time in the past \citep{kwo89}. However, the compact
hydrocarbon-dust emitting region found in this paper implies that
carbon-dust formation is still in progress in IRAS 04296+3429. This
provides a direct proof that aromatic and aliphatic hydrocarbon dust
are being made in the circumstellar environment during the post-AGB
phase.

Since the distribution of the hydrocarbon dust emission has been
measured in three PPNs, we can use these results to determine their
evolutionary status within the post-AGB evolution sequence. The wider
FWHM compared to the continuum emission region is interpreted as the
hydrocarbon dust being spatially extended. Its total luminosity is at
least comparable to that of the continuum emission.
The hydrocarbon dust around IRAS 22272+5435 is distributed between 700
and 1500~AU from the central star \citep{got03}. This is twice larger
than the hydrocarbon-dust core in IRAS 04296+3429. We applied the same
analysis to IRAS 22272+5435 as to IRAS 04296+3429 to see if there is
any extended but compact emission core in the central region of the
nebula (Fig. \ref{f3}). We found no excess of the FWHM at the central
star, although IRAS 22272+5435 is closer to us ($d =$ 1.8~kpc; private
communication, Nakashima 2006 based on the formula presented in
Deguchi et al. 2002)\footnote{The distance to IRAS 22272+5435 was
updated from 1.6~kpc previously used in Goto et al. 2003, and Ueta et
al. 2001. The physical size of the nebula is scaled up
accordingly. This new estimation is based on a simple comparison of
the apparent luminosity of the nebula obtained from photometric data
ranging from the visible to the infrared, with an assumed intrinsic
luminosity of 8,000~$L_\odot$, typical for a PPN. The interstellar
extinction is not accounted for, which may affect the visible part of
the SED. This distance is therefore an upper limit, yielding an upper
limit for the physical extent of the hydrocarbon emitting region at
IRAS 22272+5435.}, indicating that the hydrocarbon dust in IRAS
22272+5435 has been cleared up and the central star is already seen
through the dust envelope.

No enhancement of the width of the spatial profile is observed at the
hydrocarbon emission wavelength for IRAS 05341+0852 (Fig.~\ref{f3}).
At first glance, the FWHM spectrum of IRAS 05341+0852 looks similar to
that of IRAS 22272+5435 where the dust emitting region extends to as
much as 1\farcs5 in diameter. However, we demonstrate here that in
fact the hydrocarbon emission in IRAS 05341+0852 is compact and
unresolved.  First, a crucial difference between the two sources is
the lack of a diffuse hydrocarbon dust emission around IRAS
05341+0852. We ruled out any possible spatial variation of the
hydrocarbon emission feature in IRAS 05341+0852 using the same
analysis as for IRAS 22272+5435 in \citet{got03}. If positively
detected, a spatial variation would have provided unambiguous evidence
of a physically extended hydrocarbon dust emitting zone. However, not
only was such a variation not detected, but even the intensity of the
emission feature relative to the continuum level is constant
regardless of the distance from the central star, which is consistent
with a point source and favors the interpretation that the hydrocarbon
emission in IRAS 05341+0852 is unresolved. Second, we estimated the
shape of dust emitting region of IRAS 05341+0852, assuming a general
morphology similar to that of IRAS 22272+5435. The nebulosity around
IRAS 22272+5435 spans 3\farcs47$\times$3\farcs37 in {\it I}-band
\citep{uet00}, and is smaller in the mid-infrared
(1\farcs3$\times$1\farcs0 at 8~$\mu$m, averaged from Meixner et
al. 1997 and Ueta et al. 2001). This is roughly the dimension of the
detached hydrocarbon emitting zone located by \citet{got03}. IRAS
05341+0852 appears much smaller in {\it I}-band, presumably because it
is further (1\farcs14$\times$0\farcs78, Ueta et al. 2000). If we
simply scale the near-infrared appearance of IRAS 05341+0852 in
proportion to the dimension of IRAS 22272+5435 in {\it I}-band and the
mid-infrared, the hydrocarbon emitting region of IRAS 05341+0852
should measure 0\farcs2--0\farcs4 from the central star. If IRAS
05341+0852 had a detached hydrocarbon emitting region comparable to
that of IRAS 22272+5435, it would have been spatially resolved by our
observation, even when accouting for the increaded distance to IRAS
05341+0852.


Here we try to place an upper limit on the physical extent of the
hydrocarbon dust in IRAS 05341+0852. The distance to the PPN is poorly
known. We take a conservative value of $d=$ 7.4~kpc to IRAS 05341+0852
from \citet{fuj02}, which assumes the typical intrinsic luminosity of
PPNs to be $L_\ast = 8,000~L_\odot$. The FWHM of IRAS 05341+0852 is
averaged to 174.5$\pm$8.5~mas in the range of 3.2 to 3.6~$\mu$m. The
uncertainty sets upper limit for the physical extent of hydrocarbon
emitting zone to 54.2~mas after deconvolution, which translates to
400~AU at the distance quoted above. Therefore, if IRAS 05341+0852 has
a dust-emission core with similar spatial extent as IRAS 04296+3429
($\sim$400~AU), it should have been detected even at that distance,
although admittedly only marginally. In order to be consistent, if we
use 7.1~kpc as the distance to IRAS 04296+3429 as proposed by the same
authors using the same method, instead of the 4~kpc given by
\citet{mei97}, the physical extent of the hydrocarbon emitting zone
around IRAS 04296+3429 increases to $D\approx$ 1,100~AU. This makes it
even safer to say that the physical extent of IRAS 05341+0852 is less
than that of IRAS 04296+3429, and if the former had a hydrocarbon core
similar to the other one in size, it would be detectable. We therefore
conclude IRAS 05341+0852 is unresolved not because of its remote
location, but because the hydrocarbon emission region is intrinsically
small in size.

We now have the distribution of hydrocarbon dust in three PPNs: the
emitting region in IRAS 05341+0852 is unresolved; in IRAS 04296+3429
it is concentrated but extended up to 400--640~AU in diameter; and in
IRAS 22272+5435 it is diffuse and widely developed over 2000~AU. This
is at least suggestive that there is an evolutionary sequence of the
hydrocarbon production in these PPNs: IRAS 05341+0852 is in the
earliest stage of the evolution, followed by IRAS 04296+3429, with
IRAS 22272+5435 being the most evolved in the post-AGB evolution.

Although the present sample is small, we do note that there seems to
be a correlation between the size of the emitting region and the
spectral evolution of the hydrocarbon emission features. The two
trends in the spectral sequence, (1) the decline of aliphatic emission
at 3.4~$\mu$m with respect to the aromatic emission at 3.3~$\mu$m, and
(2) the decrease of the overall intensity of hydrocarbon feature
integrated at 3.2--3.6~$\mu$m, apparently go together with the
hydrocarbon emitting region becoming spatially more developed.
It is interesting to note that a similar spectral variation of the
aliphatic and aromatic hydrocarbon features has been reported in the
thermal annealing of the laboratory analogs of carbon dust
\citep{sch99,got00}.
These results show that high spatial resolution infrared imaging
spectroscopy has the capability of directly tracing the chemical
evolution of circumstellar synthesis of carbonaceous compounds, with
significant implications on our understanding of the chemical
enrichment of the interstellar medium and the Galaxy.

\acknowledgments

We thank all the staff of the Subaru Telescope and NAOJ for their
invaluable assistance in obtaining these data and for their continuous
support during IRCS and Subaru AO construction. MG is supported by a
Japan Society for the Promotion of Science fellowship.


\begin{deluxetable}{lllcrrcccl}
\tabletypesize{\scriptsize}
\tablecaption{\label{tb1}}
\tablewidth{0pt}
\tablehead{
\colhead{                    } &
\colhead{R.A. (J2000)        } &
\colhead{Dec. (J2000)        } &
\colhead{Spectral type       } &
\colhead{$V$ mag             } &
\colhead{$L$ mag             } &
\colhead{$d$ [kpc]           } &
\colhead{$L_\ast$ [$L_\odot$]} &
\colhead{$T_d$ [K]           } &
\colhead{References          }
}
\startdata
IRAS 04296+3429 & 04 32 57.0 & +34 36 13 & G0Ia  &14.17 &7.4  & 4.0/7.1
& 5300/(8000\tablenotemark{a}) & 200 & 1,2,3,4  \\
IRAS 05341+0852 & 05 36 55.0 & +08 54 08 & F4Iab &13.61 &8.13 & 7.4
    & (8000\tablenotemark{a})      & 210 & 1,2,5    \\
IRAS 22272+5435 & 22 29 10.4 & +54 51 07 & G5Ia  & 9.10 &4.43 & 1.8
    & 13000                        & 202 & 1,6,7,8
\enddata
\tablerefs{
(1) Ueta et al. 2000;
(2) Fujii et al. 2002;
(3) Manchado et al. 1989;
(4) Meixner et al. 1997;
(5) Geballe \& van der Veen 1990;
(6) Hrivnak \& Kwok 1991;
(7) Nakashima 2006;
(8) Ueta et al. 2001
}
\tablenotetext{a}{The total luminosity is assumed to be $L_\ast =
    8,000L_\odot$ when estimating the distance to the source.}
\end{deluxetable}


\begin{thebibliography}{}




\bibitem[Cl\'ement et al.(2003)]{cle03} Cl\'ement, D., Mutschke, H.,
     Klein, R., \& Henning, Th. 2003, \apj, 594, 642

\bibitem[Deguchi et al.(2002)]{deg02}
      Deguchi, S., Fujii, T., Nakashima, J., \& Wood, Peter R. 2002,
      \pasj, 54, 719


\bibitem[Ehrenfreund \& Charnley(2000)]{ehr00}
     Ehrenfreund, P.,  \& Charnley, S. B. 2000, \araa, 38, 427

\bibitem[Fujii, Nakada, \& Parthasarathy(2002)]{fuj02}
     Fujii, T., Nakada, Y., \& Parthasarathy, M. 2002, \aap, 385, 884

\bibitem[Gaessler et al.(2002)]{gae02}
     Gaessler, W., et al. 2002, Proc. SPIE, 4494, 30

\bibitem[Geballe(1997)]{geb97}
     Geballe, T. R. 1997, in ASP Conf. Ser. 122, From Stardust to
     Planetesimals, ed. Y. J. Pendleton \& A. G. G. M. Tielens (San
     Francisco: ASP), 119

\bibitem[Geballe et al.(1992)]{geb92}
     Geballe, T. R., Tielens, A. G. G. M., Kwok, S., \& Hrivnak,
     B. J. 1992, \apjl, 398, L89

\bibitem[Geballe \& van der Veen(1990)]{geb90}
     Geballe, T. R., \& Van der Veen, W. E. C. J. 1990, \aap, 235, L9


\bibitem[Goto et al.(2003)]{got03}
     Goto, M., Gaessler, W., Hayano, Y., Iye, M., Kamata, Y., Kanzawa,
     T., Kobayashi, N., Minowa, Y., Saint-Jacques, D. J., Takami, H.,
     Takato, N., \& Terada, H. 2003, \apj, 589


\bibitem[Goto et al.(2000)]{got00}
     Goto, M., Maihara, T., Terada, H., Kaito, C., Kimura, S., \& Wada,
     S. 2000, \aaps, 141, 149.


\bibitem[Henning \& Mutschke(2001)]{hen01}
    Henning, Th.,  \&  Mutschke, H.\ 2001, Spectrochimica Acta, 57, 815

\bibitem[Henning, J\"ager, \& Mutschke(2004)]{hen03}
      Henning, Th., J\"ager, C., \& Mutschke, H. 2003, ASP
      Conf.~Ser.~309: Astrophysics of Dust, 309, 603

\bibitem[Henning \& Salama(1998)]{hen98}
      Henning, T., \& Salama, F. 1998, Science, 282, 2204

\bibitem[Hrivnak \& Kwok(1991)]{hri91}
      Hrivnak, B.~J., \& Kwok, S.\ 1991, \apj, 368, 564

\bibitem[Kobayashi et al.(2000)]{kob00}
      Kobayashi, N., et al. 2000, Proc. SPIE, 4008, 1056

\bibitem[Kwok(1993)]{kwo93}
     Kwok, S. 1993, \araa, 31, 63

(Cambridge: Cambridge University Press)

\bibitem[Kwok(2004)]{kwo04}
     Kwok, S. 2004, Nature, 430, 985

\bibitem[Kwok, Hrivnak, \& Geballe(1995)]{khg95}
     Kwok, S., Hrivnak, B.J., \& Geballe, T.R. 1995, \apj, 454, 394

\bibitem[Kwok, Volk, \& Bernath(2001)]{kwo01}
     Kwok, S., Volk, K., \& Bernath, P. 2001, \apjl, 554, L87

\bibitem[Kwok, Volk, \& Hrivnak(1989)]{kwo89}
Kwok, S., Volk, K. M., \& Hrivnak, B. J. 1989, \apjl, 345, 51

\bibitem[Kwok, Volk, \& Hrivnak(1999)]{kwo99}
     Kwok, S., Volk, K., \& Hrivnak, B. J. 1999, \aap, 350, L35


\bibitem[Lord(1992)]{lor92}
      Lord, S. D. 1992, A New Software Tool for Computing Earth's
      Atmosphere Transmissions of Near- and Far-Infrared Radiation, NASA
      Technical Memoir 103957 (Moffett Field, CA: NASA Ames Research
      Center)

\bibitem[Manchado et al.(1989)]{man89}
     Manchado, A., Garcia-Lario, P., Esteban, C., Mampaso, A., \&
     Pottasch, S.~R.\ 1989, \aap, 214, 139

\bibitem[Meixner et al.(1997)]{mei97}
     Meixner, M., Skinner, C. J., Graham, J. R., Keto, E., Jernigan,
     J. G., \& Arens, J. F.  1997, \apj, 482, 897


\bibitem[Papoular et al.(1996)]{pap96}
     Papoular, R., Conard, J., Guillois, O., Nenner, I., Reynaud, C., \&
     Rouzaud, J.-N. 1996, \aap, 315, 222

\bibitem[Pendleton \& Allamandola(2002)]{pen02}
     Pendleton, Y. J., \&  Allamandola, L. J. 2002, \apjs, 138, 75


\bibitem[Sahai (1999)]{sah99}
     Sahai, R. 1999, \apjl, 524, L125

\bibitem[Schnaiter et al.(1999)]{sch99}
     Schnaiter, M., Henning, Th., Mutschke, H., Kohn, B., Ehbrecht, M.,
     \& Huisken, F. 1999, \apj, 519, 687


\bibitem[Takami et al.(2004)]{tak04}
     Takami, H., et al. 2004, \pasj, 56, 225



\bibitem[Tokunaga et al.(1998)]{tok98}
     Tokunaga, A. T., et al. 1998, Proc. SPIE, 3354, 512


\bibitem[Ueta, Meixner, \& Bobrowsky(2000)]{uet00}
     Ueta, T., Meixner, M., \& Bobrowsky, M. 2000, \apj, 528, 861

\bibitem[Ueta et al.(2001)]{uet01}
     Ueta T., et al. 2001, \apj, 557, 831

\end{thebibliography}
\end{document}